# Rotation angle sensing via polarization-dependent mode superposition in a hollow-core fiber with two-fold symmetry

Enzo Dantas da Silva [a], Giovanna M. R. Castro [a], Murilo L. Batista [a], Flavio A. M. Marques [a], Alexandre A. C. Cotta [a], Jefferson E. Tsuchida [a], Julio C. Ugucioni [a], Leomar S. Marques [a], Matthieu Chafer [b], Benoît Debord [b], Frédéric Gérôme [b], Fetah Benabid [b], Jonas H. Osório [a, *]

[a] *Multiuser Laboratory of Optics and Photonics, LaMOF, Department of Physics, Federal University of Lavras, Lavras, Brazil*
[b] *GPPMM Group, XLIM Institute, University of Limoges, UMR CNRS 7252, Limoges, France*

**Abstract**

Hollow-core photonic crystal fibers (HCPCF) have experienced tremendous advancements recently, leading to remarkable demonstrations of transmission loss reduction and deeper understanding of their guidance fundamentals. Indeed, this progress has entailed investigations into various HCPCF designs, allowing for attaining diverse properties of interest such as ultralow loss, polarization filtering, and specific modal operation. Among HCPCFs with tailored modal characteristics are fibers displaying microstructures with modified symmetry, which allow for changing the loss hierarchy between the guided modes, hence favoring the propagation of higher-order modes. In this context, we here demonstrate the realization of an angle sensor utilizing a tubular HCPCF with a two-fold symmetric cladding. This specific fiber design enables the generation of an output intensity profile resulting from the superposition of $LP_{01}$ and $LP_{11}$-like modes, whose excitation and resulting output intensity spatial distribution are dependent on the polarization angle of the incident light. Thus, by rotating the input beam's polarization and analyzing the evolution of the resulting output profile, we characterized a rotation angle sensor exhibiting a sensitivity of $(25 \pm 1)$ counts/degree and an estimated resolution of 0.3º. We understand that this work broadens the framework of HCPCF applications, demonstrating that symmetry-modified hollow-core fibers can act as a promising platform for advanced sensing scenarios and polarimetric characterizations.

## 1. Introduction

Hollow-core photonic crystal fiber (HCPCF) technology has witnessed impressive advancements in recent years, establishing itself as a robust platform for addressing diverse application needs. Indeed, significant efforts have been dedicated by the HCPCF community toward optimizing fiber designs and fabrication processes, particularly aimed at minimizing transmission loss values. This family of optical fibers currently provides the lowest loss figures across all fiber optics guiding from the ultraviolet to the infrared [1-3].

In this context, understanding and tailoring the modal properties of HCPCFs is paramount for optimizing their characteristics, such as achieving low loss and broad transmission bandwidth, and controlling their effective single-mode or multimode behavior [4-10]. For example, by adequately choosing the ratio between the sizes of the cladding elements and core diameters, the fiber structure can be designed to resonantly filter higher-order modes and attain effective single-mode propagation [4-6]. Additionally, large-core HCPCFs can be used to explore their multimode behavior in beam delivery and sensing applications [7-10]. Moreover, altering the fiber cladding symmetry properties allows for achieving specific modal characteristics. For instance, shifting the azimuthal position of the tubes in the cladding allows for rearranging the loss hierarchy of the modes and yield salient polarization properties [11]. This approach, according to the fiber design explored, allows for the preferential propagation of higher-order core modes, such as $LP_{11}$ and $LP_{21}$-like modes over the fundamental mode [11].

Additionally, the strategic modification of the cladding enables the controlled excitation of combinations of core modes within the fiber structure. This capability, coupled with the distinct polarization properties of these modes, allows for the generation of complex output intensity and polarization distributions. Such fibers can therefore function as mode and polarization shapers [11]. Noteworthily, this concept can be explored in a variety of applications, including atom optics, atom-surface physics, sensing, and nonlinear optics [11-13].

Considering this scenario, in this work, we demonstrate a novel approach to perform rotation angle sensing by utilizing a HCPCF engineered with a two-fold symmetric cladding microstructure. This specific fiber design, which enables the generation of an output intensity profile resulting from the superposition of $LP_{01}$ and $LP_{11}$-like modes, is such that the spatial distribution of its output beam is dependent on the polarization angle of the incident beam. Thus, by systematically rotating the input beam's polarization angle and analyzing the resulting fiber output profile, we characterized a rotation angle sensor exhibiting a sensitivity of $(25 \pm 1)$ counts/degree and an estimated resolution of 0.3º. Therefore, we understand that this work broadens the applications of HCPCFs within the sensing area, particularly those encompassing fibers displaying microstructures with modified symmetry and polarimetric devices.

## 2. Fiber structure and its output intensity profile

The fiber employed in this research is a tubular HCPCF displaying a two-fold symmetry cladding. Its cross-section, shown in Fig. 1, is characterized by a structured cladding formed by 8 tubes which are unevenly distributed around the hollow core, in such a way that two larger gaps are obtained. The fiber possesses a core diameter of 45 μm. Its cladding tubes have an outer diameter of 15 μm and a thickness of 750 nm. The larger gaps between the cladding tubes measure 17 μm.

As demonstrated in [11], the strategic modification of the azimuthal position of the cladding tubes and the corresponding enlargement of the gaps between them alter the modes' loss hierarchy, leading to the preferential propagation of higher-order core modes. This contrasts with typical, fully symmetric fibers in which the fundamental mode usually has the lowest loss. Indeed, this special two-fold symmetric microstructure of the cladding allows for obtaining a polarization-dependent intensity profile at the fiber output resulting from the superposition of $LP_{01}$ and $LP_{11}$-like modes, which are the lowest loss core modes supported by this fiber structure [11]. This hence enables this fiber to act as an angle sensor, as we will describe in the following.

To qualitatively describe the output profile of the fiber, which is dominated by the superposition of its lowest loss, $LP_{01}$ and $LP_{11}$-like modes, one can firstly write the corresponding electric field components in the vertical and horizontal polarizations, $E_X$ and $E_Y$, respectively, as Eq. (1) and (2). Here, $\alpha$, $\beta$, $\gamma$, and $\delta$ are constants accounting for the contributions of the respective modes to the output profile. In turn, the resultant output intensity profile, $|E_R(\theta)|^2$, can be expressed by Eq. (3), where $\theta$ stands for the polarization angle of a linearly polarized input beam coupled to the fiber [11]. As we will propose in the next paragraphs, the fiber output intensity profile, being dependent on the polarization characteristics of the incident beam, allows for defining a metric to correlate angle variations with the corresponding optical response of the system.

$$E_X = \alpha\, E_{LP01,X} + \beta\, E_{LP11,X} \qquad (1)$$

$$E_Y = \gamma\, E_{LP01,Y} + \delta\, E_{LP11,Y} \qquad (2)$$

$$|E_R(\theta)|^2 = (E_X \cos\theta)^2 + (E_Y \sin\theta)^2 \qquad (3)$$

To illustrate the operational principle of our angle sensor, we conducted a simplified analytical analysis to qualitatively describe the fiber output intensity profiles as a function of the input polarization angle. To do so, the electric field components of the fundamental $LP_{01}$ mode have been modeled using $E_{LP01} = J_0(u_{01}x/a)$ and the $LP_{11}$-like mode components have been modeled using $E_{LP11} = J_1(u_{11}x/a)$, where $J_n$ is the n-th order Bessel function of the first kind, $u_{01}$ is the first zero of $J_0$, $u_{11}$ is the first zero of $J_1$, and $a$ is the core radius. The resulting intensity profile is calculated using Eq. (3). Here, we mention that, although these expressions do not reproduce the true modes in the studied fiber structure, they allow us to visualize the changes of the spatial distribution of the optical beam at the fiber output as the input polarization angle is altered, hence contributing to the understanding of the sensing platform working principle.

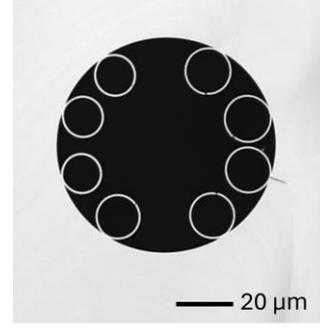

Fig. 1 Image of the cross-section of the HCPCF with two-fold symmetry used in this work. The fiber cladding is formed by 8 tubes and displays two larger gaps between each group of four tubes. The fiber core has a 45 µm diameter, and the cladding tubes has a 15 µm outer diameter and thickness of 750 nm. The larger gaps between the cladding tubes measure 17 µm.

The fundamental mechanism enabling angle sensing is the polarization-dependent interference between the supported core modes, as detailed in the modal decomposition shown in Fig. 2. Fig. 2a and 2b depict the normalized electric field distributions for the horizontal and vertical polarization components of $LP_{01}$ and $LP_{11}$ modes (blue and red lines, respectively), and the corresponding resultant intensity profile that arises from the in-phase superposition of both (purple lines), considering equal contributions of the $LP_{01}$ and $LP_{11}$ modes in the fiber modal content (i.e., $\alpha = \beta = \gamma = \delta = 0.5$). For an input polarization angle of $\theta = 0°$, the interference between $LP_{01}$ and $LP_{11}$ results in a localized intensity peak on the negative side of the transverse axis. Conversely, at $\theta = 90°$, the excitation of vertical components causes the intensity peak to appear on the positive side of the transverse axis. As we will see in the following the transition from these two intensity profiles can be followed as $\theta$ is varied, hence enabling angle sensing capabilities in our platform.

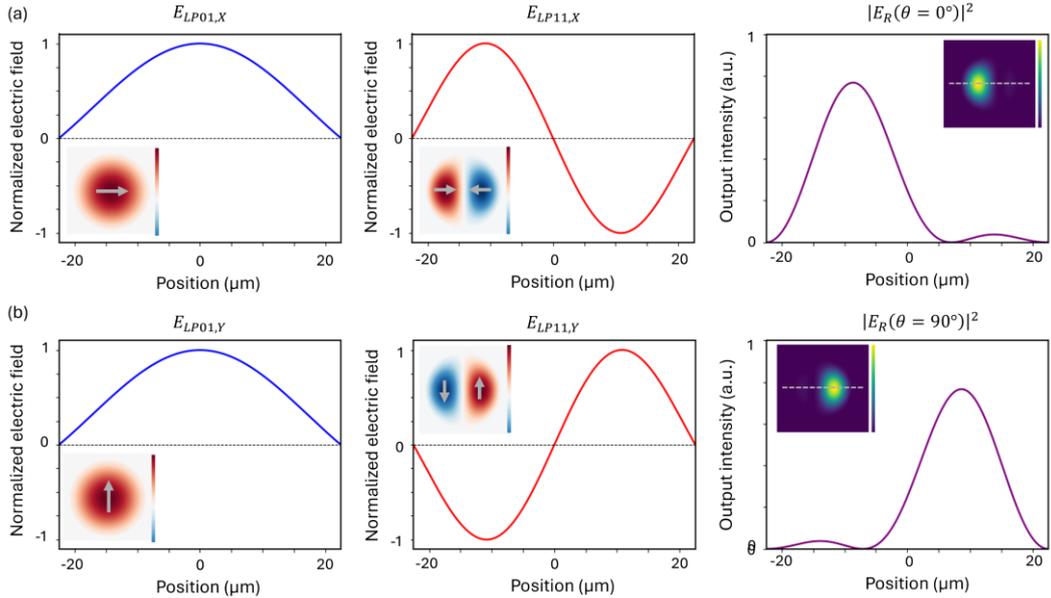

Fig. 2 Analytical modeling of the polarization-dependent modal interference at the fiber output. Normalized electric field distributions and corresponding intensity profiles are shown for the (a) horizontal and (b) vertical polarization components. The analysis considers an in-phase interference condition between the $LP_{01}$ and $LP_{11}$-like modes. The insets illustrate the electric field distribution, the polarization orientation of the modes and the intensity spatial distribution resulting from the superposition between the modes for $\theta = 0°$ and $\theta = 90°$.

To further elucidate this transduction principle, Fig. 3 presents the evolution of the fiber output intensity profile as a function of the input polarization angle $\theta$, varied from 0° to 90° in 5° increments, calculated by considering the model described above. The 3D waterfall plot shown in Fig. 3 readily illustrates the spatial redistribution of power across the fiber core as a function of the variation of the input polarization angle. As $\theta$ increases from 0°, the intensity of the left-hand lobe gradually diminishes while the right-hand lobe strengthens (attaining a maximum when $\theta = 90°$). At the intermediate angle of $\theta = 45°$, the contributions from both polarization states are balanced, resulting in a dual-peak distribution. The variation in the beam spatial distribution relative to the polarization angle hence allows for determining a metric for angle sensing, enabling the two-fold symmetric HCPCF to act as a geometric phase-like converter that maps polarization state to spatial intensity distribution.

### 3. Fiber output experimental characterization

To validate the proposed sensing principle and characterize the platform acting as a rotation angle sensor, we employed the setup depicted in Fig. 4. The experimental setup encompasses a linearly polarized He-Ne laser source emitting at 633 nm, alignment mirrors (M1 and M2), a half-wave plate ($\lambda/2$), input and output lenses (L1 focuses the input beam in the HCPCF and L2 collimates the output beam towards the measuring system), and a CMOS camera. For the experimental results reported herein, we employed a HCPCF with a length of 3 m to achieve the modal superposition required for the sensing measurements. It should be noted, however, that the system is adaptable to alternative fiber lengths, provided the input coupling is optimized to maintain the desired output intensity distribution. During the sensing measurements, the near-field intensity profile at the fiber output has been measured by the CMOS camera as a function of the waveplate rotation angle, $\alpha$.

The top row of Fig. 5 displays representative experimental images of the fiber output beam for selected waveplate rotation angles, namely $\alpha = 0°$, 20° and 40° (or, in terms of the polarization angle, $\theta = 0°$, 40° and 80°). At $\alpha = 0°$, the intensity distribution is predominantly concentrated in the left side of the core, corresponding to the constructive interference of the x-polarized components of the $LP_{01}$ and $LP_{11}$-like modes, as expected by the analytical description presented in the previous section. As the waveplate is rotated to $\alpha = 20°$ ($\theta = 40°$), a spatial redistribution of the intensity profile is observed, with a second lobe emerging on the right side of the core. At $\alpha = 40°$ ($\theta = 80°$), the spatial flip is nearly complete, with the peak intensity now localized on the right-hand side of the core.

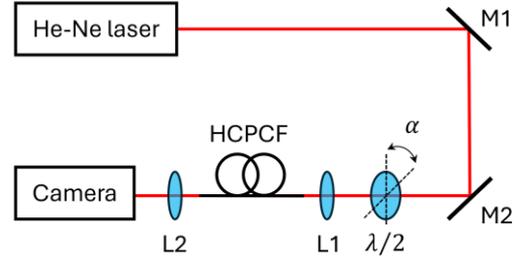

Fig. 4 Diagram of the experimental setup for the rotation angle sensing measurements. M1 and M2: alignment mirrors; L1 and L2: lenses; $\lambda/2$: half-wave plate; $\alpha$: waveplate rotation angle.

The evolution of this spatial redistribution is further elucidated in the 3D waterfall plot shown in Fig. 5. Each near-field transverse profile measured by the CMOS camera is accounted for along the horizontal axis of the fiber, as indicated by the dashed lines in Fig. 5, and corresponds to a specific waveplate rotation angle. Observation of the intensity profiles shown in Fig. 5 shows that the experimental data is in qualitative agreement with the analytical model presented in Section 2. Specifically, the transition of the beam's intensity spatial distribution across the core confirms that the two-fold symmetric HCPCF microstructure allows for determining a metric to account for the variation of incident polarization angle, which directly correlates with the rotation angle of the waveplate. We will discuss this metric in the following section.

### 4. Rotation angle sensing

The angular sensing performance of the two-fold symmetric HCPCF studied herein was characterized by monitoring the peak intensity evolution of the two symmetric interference lobes at the fiber output as the rotation angle of the waveplate was varied. To capture this behavior, two regions of interest (ROIs) were defined, within which the intensity-weighted center of mass, or centroid, was calculated for each $\alpha$. The two ROIs have been defined as ROI 1, corresponding to the left side of the core (pixel positions from -225 to -75), and ROI 2, corresponding to the right side of the core (pixel positions from 25 to 175). Fig. 6a indicates ROI1 (in blue) and ROI2 (in pink). The black line stands for the profile measured for $\alpha = 0°$.

As shown in Fig. 6b, the peak intensity in ROI 1 displays a decreasing trend as $\alpha$ is rotated from 0° to 20°. In turn, the peak intensity in ROI 2 exhibits an increasing trend for $20° < \alpha \leq 40°$, with a crossover point corresponding to the data in ROI1 occurring around 20°. To transform these dual-peak evolutions into a unified sensing metric, a piecewise linear calibration was implemented. A zero-reference level was established by averaging the centroid intensities of both lobes at the 20° crossover point. This reference level was then subtracted from the peak intensity data for both regions, leading to corrected intensity values. The calibration curve was then constructed by selecting the corrected intensities of ROI 1 for the $0° \leq \alpha \leq 20°$ range and ROI 2 for the $20° < \alpha \leq 40°$ range, resulting in a rising slope as shown in Fig. 6c. We mention that, to maintain such a unified rising slope, the first segment (ROI 1) was mathematically inverted by applying a negative sign to its corrected values. This procedure stitched the decreasing behavior of the left lobe to the increasing behavior of the right lobe, hence eliminating the non-linear behavior that would otherwise occur if the lobes were monitored individually.

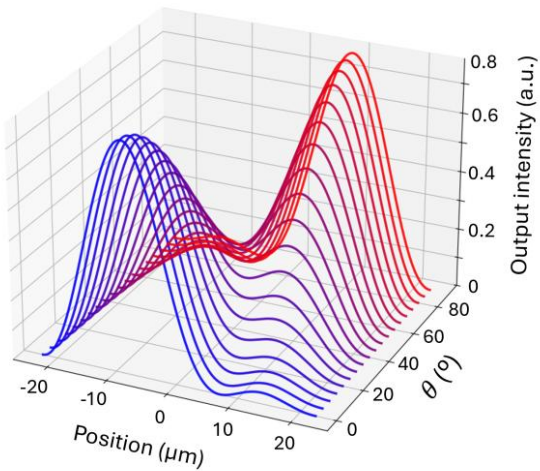

Fig. 3 3D waterfall plot showing the calculated spatial redistribution of the intensity distribution at the fiber output as a function of the input beam polarization angle, $\theta$, varied from 0° to 90° in 5° increments.

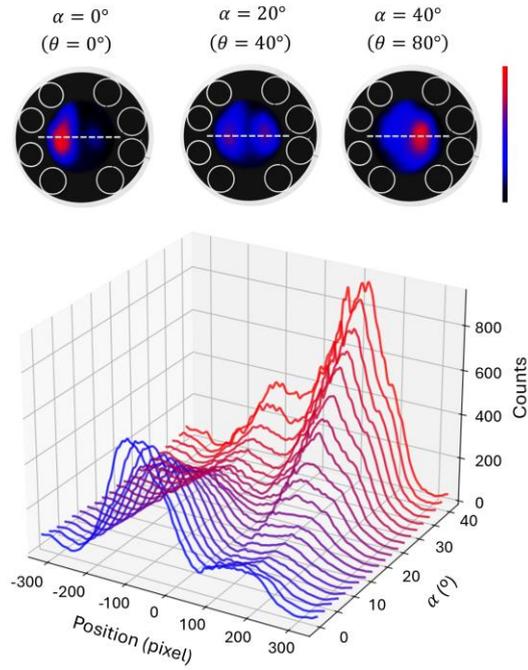

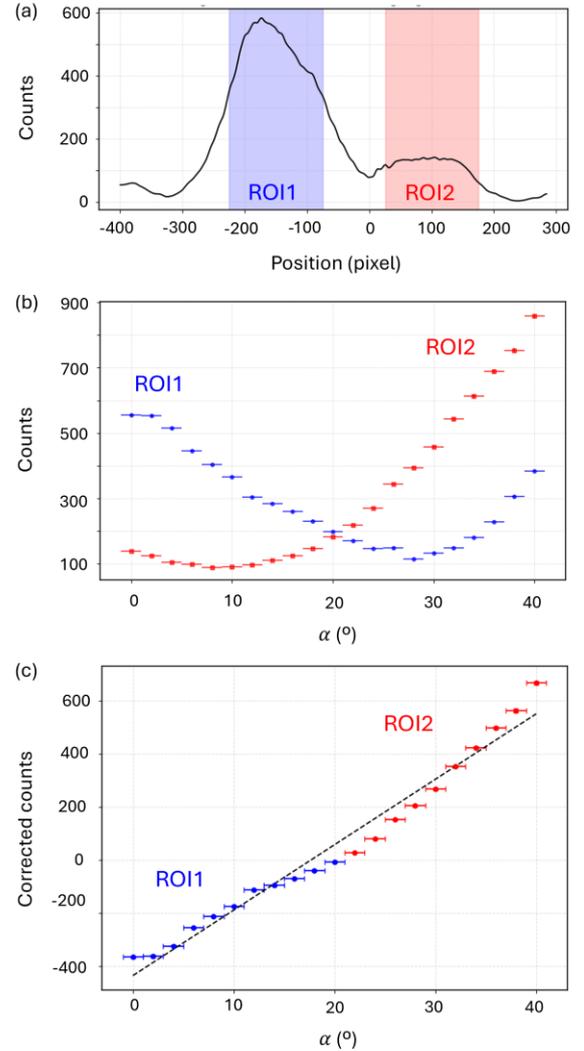

Fig. 5 3D waterfall plot showing the experimentally measured intensity profiles at the fiber output as a function of the half-wave plate rotation angle, $\alpha$. The images on the top stand for measured near-field profiles at representative $\alpha$ values. The dashed line indicates the axis from which the profiles shown in the 3D waterfall plot has been accounted.

The resulting plot, displayed in Fig. 6c, allows for defining the system calibration curve. Thus, the sensor's sensitivity was determined by accounting for the slope of the linear regression applied to this unified dataset, yielding a value of $(25 \pm 1)$ counts/degrees with a coefficient of determination of 0.9718. This methodology hence demonstrates that the assessment of the spatial redistribution of light within the core of the HCPCF with a two-fold symmetric cladding can be effectively translated into a linear metric for the realization of angular sensing.

Additionally, we mention that the error bars associated with the intensity values in Fig. 6c (which are not visible due to the axis scale) were derived from the positional uncertainty of the centroid, as determined through the fitting procedure. Across the measured range, the maximum estimated intensity error was found to be 7 counts. By treating this maximum uncertainty as the threshold for the minimum resolvable intensity change, the angular resolution of the system is estimated to be 0.3°. Indeed, the rotation angle sensor presented in this work offers an alternative to conventional fiber-based angular sensing technologies, such as those based on fiber Bragg gratings, long-period gratings, or more complex interferometers [14] as, while traditional sensors often require more sophisticated spectral analysis (besides more expensive instrumentation) to correlate wavelength shifts with angular rotation, our platform utilizes a direct spatial-intensity transduction mechanism. Therefore, we understand that the proposed platform demonstrates a promising approach for angular sensing and widens the framework of the utilization of modal interference in HCPCF with modified claddings for sensing purposes.

Additionally, we observe that, while the polarization angle variation in this study was mechanically controlled using a rotating half-wave plate, the proposed platform could be adapted to a wider range of other transduction scenarios. For example, by replacing the wave plate with an electro-optic modulator the system could be reconfigured to work as an electric field or voltage sensor [15], in which variations in the applied field would be mapped to spatial intensity variations at the fiber output. Furthermore, the sensor could be integrated into setups for monitoring processes that induce birefringence or optical activity, such as glucose concentration sensing or the detection of chiral molecules [16]. Thus, we believe that the platform we reported here provides a versatile framework for sensing parameters whose variations can be encoded into light polarization states. This versatility hence indicates this platform as a promising tool for the realization of advanced polarimetric characterization in a wide range of experimental needs.

Fig. 6 Experimental characterization of the HCPCF rotation angle sensor. (a) Representative intensity profile at $\alpha = 0°$ illustrating the definition of the two regions of interest (ROI 1 and ROI 2) used for centroid peak tracking. (b) Individual evolution of the centroid peak intensities for ROI 1 and ROI 2 as a function of the waveplate polarization angle. (c) Unified piecewise calibration curve. The linear regression (dashed line) yields a sensitivity of $(25 \pm 1)$ counts/degrees.

## 5. Conclusions

In this work, we demonstrated the realization of a rotation angle sensor based on a tubular HCPCF with a two-fold symmetric cladding microstructure. This special fiber architecture enables the excitation and superposition of $LP_{01}$ and $LP_{11}$-like modes, resulting in an output intensity spatial distribution that is intrinsically dependent on the polarization angle of the incident light beam. The study commenced by analyzing the sensing platform's operational principles through a simplified analytical description, which was subsequently followed by the experimental demonstration of the device. By systematically rotating the input beam's polarization and analyzing the evolution of the resulting peak intensities in the fiber output beam profile via centroid tracking, we characterized a platform that maps the rotation of the incident polarization (correlated here to the rotation angle of a half-wave plate) to specific spatial intensity distributions. These experimental results confirmed that the spatial redistribution of optical power within the fiber core can be simplified into a unified linear metric through a piecewise calibration procedure. Specifically, the system demonstrated a sensitivity of 25 counts/degree with an estimated resolution of 0.3°. Ultimately, the results reported herein demonstrate that symmetry-modified HCPCFs can act as a promising platform for advanced polarimetric characterization and sensing applications, thereby broadening the framework of hollow-core fiber technology within sensing scenarios.

## Acknowledgments

The authors thank FAPEMIG (RED-00046-23, APQ-00197-24, APQ-01401-22, APQ-01618-25, APQ-01765-22, BIS-00429-25) and CNPq (305024/2023-0, 402723/2024-4, 409174/2024-6, 447347/2025-0) for financial support. The authors also thank the support from the National Institute of Photonics (INFO), Brazil, and Finep.